\documentclass[twocolumn,showpacs,amsmath,amssymb]{revtex4}
\usepackage{graphicx}
\usepackage{bm}
\begin{document}
\title{Magnetic field-induced gapless state in multiband superconductors.}
\author{Victor Barzykin} 
\affiliation{Department of Physics and Astronomy, University of Tennessee,
Knoxville, TN  37996-1200}

\begin{abstract}
We investigate theoretically the properties of s-wave multiband superconductors 
in the weak coupling (BCS) limit in the presence of pair-breaking effects of  magnetic field.
It is shown that a qualitatively new gapless superconducting state must appear in quasi-2D superconductors
in magnetic fields parallel to the plane, corresponding to a Sarma state induced on one of the Fermi surfaces. 
The emergence of the new state in s-wave multiband superconductors in the absence of anisotropy or spin-orbit
interaction is usually accompanied by a zero-temperature first-order metamagnetic phase transition. 
For anisotropic or non s-wave multiband superconductors the order of the zero-temperature 
metamagnetic transition depends on model parameters, and it may take the form of a smooth crossover. 
The details of the temperature-magnetic field phase diagram for multiband superconductors are
investigated analytically at zero temperature and numerically at a finite temperature. It is shown 
the zero-temperatures first-order phase transition gives rise to a critical region on the $B-T$ phase
diagram. We suggest possible experiments to detect the new gapless state.
\end{abstract}
\vspace{0.15cm}


\maketitle

\section{Introduction}  

Superconductivity in multiband metals was first investigated theoretically shortly after the BCS
theory\cite{suhl,moskalenko,geilikman,kresin,cohen}. A resurgence of interest in multiband superconductivity  
has been mostly related to experimental discovery of  MgB$_2$\cite{MgB2}, and observation of the two s-wave gaps 
by various techniques\cite{giubileo,iavarone,szabo,schmidt,wang,bouquet1,yang,bouquet2}. Recent discovery
of iron oxypnictides\cite{Kamihara}, a new family of quasi-2D high temperature superconductors, has also attracted
enormous theoretical interest to the problem of multiband superconductivity. First principles
numerical band structure calculations\cite{nekrasov,DJSingh,MazinL} show  that several bands cross
the Fermi level in these materials.
While the pairing mechanism in iron oxypnictides is not yet clear, several bands 
are involved in the determination of both normal and superconducting properties\cite{MazinL,BG1}. 
The existence of multiband energy spectrum\cite{mazin} and associated several energy
gaps\cite{yokoya,fletcher} was also verified experimentally in an older, quasi-2D s-wave superconductor
NbSe$_2$.  BCS investigations of multiband superconductivity were restarted\cite{zhitomirsky,koshelev},
but were mostly centered around the new physics that arises due to the presence of two separate gaps.
The multiband energy spectrum is also present in most unconventional heavy fermion superconductors due to extreme
complexity of the band structure of these materials, as indicated by de Haas van Alfven measurements. For example,
gapless superconductivity of Abrikosov-Gor'kov type\cite{AG} was recently observed
in thermal conductivity data for La-doped CeCoIn$_5$\cite{tanatar}, 
as indicated by  unusual Wiedeman-Franz $1/x$ behavior of thermal conductivity at small concentration
of La, $x$. The unusual behavior was attributed to the multiband structure\cite{tanatar,BG0} of this material,
a d-wave superconductor.

Theoretical study of multiband superconductivity has not only been motivated by the 
above compounds. A somewhat modified multiband model is applicable to other new materials\cite{mineevIS},
where a single Fermi surface gets spin-split into two pieces with different Fermi momenta 
due to interactions, such as superconductors without center of inversion (CI)\cite{AGT}
(for example, CePt$_3$Si\cite{bauer}),  or  ferromagnetic superconductors (UGe$_2$\cite{saxena,huxley},
 ZrZn$_2$\cite{pfleiderer}, or URhGe\cite{aoki}). A particular case of CI-symmetry breaking is
two-dimensional surface superconductivity\cite{GR,BG}, where two Fermi surfaces arise as a result
of spin-orbit interaction of Rashba form\cite{rashba1,rashba2}. 

The Fermi surface will also get spin-split in exchange field, an external magnetic field without orbital effects.
Theoretically, paramagnetic pair-breaking by exchange field corresponds to the old problem of unbalanced
pairing, first studied in the early 60-s\cite{Clogston,chand,LO,FF,sarma,takada}, when the $B-T$ phase
diagram for three-dimensional materials in exchange field has first been obtained.
Orbital effects are almost always present in three dimensions, and even in very anisotropic
quasi two-dimensional (2D) materials, unless $H_{c2}$ in magnetic fields parallel to
the 2D plane close to Clogston paramagnetic limit\cite{Clogston,chand}. This is, 
perhaps, the reason why the consequences of the theory have not studied in detail experimentally.
A study of thin films in magnetic fields parallel to the surface remains, perhaps,
the most promising experimental setup for the observation of unbalanced pairing in superconductors\cite{film1,film2,film3}.
A possible observation of the Larkin-Ovchinnikov-Fulde-Ferrel (LOFF)\cite{LO,FF} has also been reported
in some quasi 2D superconductors, such as those based on charge-transfer organic salts of BEDT-TTF or 
ET - ion\cite{Singleton} and the 1-1-5 family heavy fermion materials\cite{capan,KY,mizushima,kakuyanagi,mitrovic}.

While most theoretical studies of superconductivity in multiband compounds concentrated on the 
mechanism of superconductivity and the effects of several energy gaps on superconducting properties,
some recent studies\cite{review} found that a new class of superfluids could potentially arise
in these materials, one that features coexistence of fully gapped and gapless states.
For the most part, this class of states was proposed in 
Bose-Einstein condensation for different non-identical fermions condensed by an optical trap\cite{LW}, 
or in high-energy physics\cite{quarksW}, where the ``breached'' superfluid
state appears as a result of pairing between different quarks\cite{gubankova1,gubankova2} in
the asymptotically free limit.

Gapless solutions of this type were previously known, but in most cases were found to be energetically unstable. 
The first example corresponds to the second unstable solution in the unbalanced pairing problem, and is commonly referred to as
Sarma state\cite{sarma,takada}. The second solution of the gap equation for a superconductor placed in an exchange field 
was first obtained by Green's functions method by Baltensperger\cite{baltensperger} and Gor'kov and Rusinov\cite{Gor:Rus},
where it was found to be energetically unstable. The exact nature of this unstable state, and the
existence of unpaired electron- and hole- Fermi surfaces was later clarified within BCS theory by Sarma and 
Takada\cite{sarma,takada}).  The second example is pairing in a doped (unbalanced) excitonic gas, leading to an 
excitonic condensate\cite{butov1,butov2}, which is actually the same two-band
unbalanced pairing problem as one considered by Liu and Wilczek\cite{LW}. The 
solution previously found for this problem takes the form of a magnetically (ferromagnetically or antiferromagnetically) 
polarized  gapless state\cite{zhit:exc,VKR}, and is different from ``interior gap superfluid'' of Liu and Wilczek. 
Similar to Sarma state, this solution is usually unstable
with respect to formation of LOFF state or domains\cite{GM,BGE,BV}, although the stability of this magnetic solution 
has not been fully investigated as a function of masses involved.
The third example is the Sarma gapless state, which appears in the problem of s-wave pairing 
in ferromagnetic superconductors\cite{blagoev}. While it has been claimed\cite{blagoev} that Sarma state
is stabilized by the presence of ferromagnetic order, this claim has been debated later\cite{bl:comm1,bl:comm2}.
According to recent work of Liu and Wilczek, a large difference in effective masses on two Fermi surfaces tends to stabilize 
the ``breached'' superfluid  state already within the unbalanced pairing problem\cite{LW}. 

The main motivation for this paper is a detailed theoretical study of possible new gapless states 
in multiband superconductors in the presence of exchange magnetic field\cite{BG2}, or the multiband
unbalanced pairing problem. While certain similarity does exist, the multiband problem is 
really different\cite{ABG} from the unbalanced pairing problem considered in the above cases;
pairing between two different species of fermions (two different bands)  
is usually not a relevant mechanism in multiband superconductors, since
the energy difference for the two bands, $\Delta \epsilon \sim 1 eV$ is much greater than $T_c \sim 1 K$. 
Nevertheless, for a superconductor placed in an external ``exchange'' field the unbalanced pairing problem is recovered.
Surprisingly, as it was shown in Ref.\cite{BG2}, the multiband structure often leads to a 
a stabilization of unusual Sarma state on the second band in exchange fields $\mu_B B > \Delta_2$, 
where $\Delta_2$ is the energy band on the second Fermi surface.
Thus, in quasi-2D superconductors where several bands cross the Fermi surface, 
such as CeCoIn$_5$ of 2H-Nb$Se_2$, the peculiarities of the $B-T$ phase diagram may not 
limited to high magnetic fields, where the LOFF-related phenomena are observed.
New singularities and gapless states associated with the gaps on secondary Fermi surfaces
must arise in \textit{low} magnetic fields as well\cite{BG2}. They correspond
to the appearance of Sarma state\cite{sarma} on the Fermi surface with smaller gap,
one that becomes energetically stable due to the presence of the superconducting
gap on the other Fermi surface. This state is characterized\cite{sarma} by the presence of 
unpaired spin-polarized electrons near the Fermi surface of the second band, two open electron- and hole- Fermi surfaces, 
a paramagnetic magnetic moment, and a first-order phase transition that always accompanies
the appearance of this unusual state in s-wave multiband superconductors in low magnetic fields\cite{BG2}.

The paper is organized as follows. In Section II we introduce the multiband model and generalize its known solution 
in the s-wave case to include effects of gap anisotropy and non s-wave pairing symmetry. We demonstrate that 
the effective coupling constants can be eliminated in favor of the measurable parameters for multiband superconductors,
such as the superconducting transition temperature, $T_c$, and the ratio of the gap amplitudes 
and the densities of states on different Fermi surfaces. In particular, thermodynamics of multiband superconductors 
is additive; the thermodynamic potential $\Omega$ is a simple BCS sum over different bands. In Section III we consider the
problem of paramagnetic pair-breaking in multiband superconductors and show that a new gapless state 
is energetically stable in low magnetic fields in some region of model parameters. 
We investigate the stability and the magnetic properties of this new gapless state in an s-wave two-band
superconductor analytically at $T=0$, and provide the details for the $B-T$ phase diagram 
and the low temperature critical point that separates the new partially gapless state from fully
gapped state. In Section IV we present our conclusions.

\section{The multiband model in the weak-coupling limit.}
 
In a standard BCS approach\cite{AGD} the pairing interaction can always be eliminated in favor of a single energy scale, 
$T_c$, giving rise to the well-known universality of the BCS theory. The critical temperature $T_c$ is the only parameter
that determines thermodynamic, kinetic, and other properties in the weak coupling limit. For example, the superconducting gap at 
zero temperature, $\Delta(0)$,  is related to $T_c$ in s-wave superconductors by the universal law, 
$\Delta(0) = (\pi/\gamma) T_c \simeq 1.76 T_c$. Similar universality is applicable to non s-wave superconductors
as well, even though the universal ratio $\Delta(0)/T_c$ depends on the type of pairing. For d-wave
pairing, the weak coupling ratio of the maximum gap amplitude at $T=0$ to $T_c$ is 
$\Delta(0)/T_c = 2 \pi T_c/\gamma \sqrt{e} \simeq 2.14$\cite{SM}. Since a number of interaction parameters are
involved in a multiband model, there is no such universal relation between $\Delta_{\alpha}(0)$ on different Fermi
surfaces and $T_c$. Thus, $T_c$ cannot be the only parameter that describes the properties of multiband
superconductors. Geilikman, Zaitsev, and Kresin\cite{geilikman,kresin} showed using 
the method of Pokrovskii\cite{pokrovskii} that some universality is left in the weak coupling multiband model. 
First, the physical properties, such as thermodynamics, are often additive over different bands. Second, the ratios of the
gap amplitudes on different Fermi surfaces are temperature-independent in the weak coupling limit.
Third, Geilikman \textit{et al.}\cite{geilikman} found that all physical properties of a BCS multiband superconductor
can be expressed in terms of the transition temperature $T_c$, and other quantities measurable in the normal state, such as 
the ratios of densities of states on different Fermi surfaces, and the temperature-independent ratio gap amplitudes. 
The gap amplitudes themselves, however, are not universal. In this section we introduce the multiband model, review
some of the results of Geilikman \textit{et al.}\cite{geilikman} that we will use in other sections, and generalize their 
weak-coupling solution to the case of arbitrary anisotropic pairing.

The Hamiltonian for several separate Fermi surfaces has the following form (see, for example,  Ref. \cite{ABG}):
\begin{widetext}
\begin{equation}
H_{el} = \sum_{\alpha \sigma \bm{k}} \epsilon(\bm{k}) 
a^{\dagger}_{\alpha \sigma}(\bm{k}) a_{\alpha \sigma}(\bm{k}) 
+ {1 \over 2} \sum_{\bm{k},\bm{k'}} 
\sum_{\alpha \beta \sigma_{1-4}} V_{\alpha \beta \sigma_{1-4}}(\bm{k},\bm{k'}) 
a^{\dagger}_{\alpha \sigma_1}(\bm{-k}) a^{\dagger}_{\alpha \sigma_2}(\bm{k})
a_{\beta \sigma_3}(\bm{k'}) a_{\beta \sigma_4}(\bm{-k'}),
\label{elec}
\end{equation}
\end{widetext}
where  $\sigma_{1-4}$ are spin indices, $V_{\alpha \beta \sigma_{1-4}} (\bm{k},\bm{k'})$ 
corresponds to the model interaction for a pair of electrons from 
band $\beta$ with quasimomentum $\bm{k'}$ to band $\alpha$ with quasimomentum $\bm{k}$. 
The latter can be written in the following form:
\begin{equation}
V_{\alpha \beta \sigma_{1-4}}(\bm{k},\bm{k'}) = 
V_{\alpha \beta}(\bm{k},\bm{k'}) (i \sigma_y)_{\sigma_1 \sigma_2} (i \sigma_y)^{\dagger}_{\sigma_3 \sigma_4}.
\label{int}
\end{equation}
Note that the unbalanced pairing terms\cite{LW,gubankova1,gubankova2} are not present in this model
BCS formulation, since the mismatch between different Fermi surfaces is usually too large
(of the order of $\sim eV$) for these terms to be relevant. 
We consider below any possible type of the superconducting state. As usual, the interactions in the
model BCS Hamiltonian are taken in a factorized form,
\begin{equation} 
V_{\alpha \beta}(\bm{k};\bm{k}') = \chi(\varphi, \theta) V_{\alpha \beta} \chi(\varphi',\theta'),
\label{factor}
\end{equation}
where $\chi(\varphi,\theta)$ is the appropriate irreducible representation, normalized to unity:
\begin{equation}
\int \frac{d \Omega}{4 \pi}\, |\chi(\varphi, \theta)|^2 = 1.
\label{norm}
\end{equation}
 For example, for the s-wave pairing
\begin{equation}
\chi(\varphi, \theta) = 1,
\end{equation} 
while for the d-wave pairing,
\begin{equation}
\chi(\varphi, \theta) = \sqrt{2} \cos{(2 \varphi)}.
\end{equation} 

The energy gaps on each Fermi surface $\Delta_{\alpha} (\bm{k})$ 
can be easily expressed in terms of the interaction matrix  $V_{\alpha \beta}(\bm{k};\bm{k'})$ and
the anomalous Gor'kov functions  $F_{\beta}(i \omega_n, \bm{k})$:
\begin{eqnarray}
\label{gapeq}
\Delta_{\alpha}(\bm{k}) & = & - T \sum_{n, \beta , \bm{k'}} V_{\alpha \beta}(\bm{k},\bm{k'}) F_{\beta} (i \omega_n, \bm{k'})  \\
F_{\beta}(i \omega_n, \bm{k}) & = & \frac{\Delta_{\beta}(\bm{k})}{\omega_n^2 + \xi^2 + |\Delta_{\beta}(\bm{k})|^2}\,.
\end{eqnarray}
The transition temperature $T_c$ is determined by the linearized gap equation Eq.(\ref{gapeq}) for the gap
amplitude 
\begin{equation}
\Delta_{\alpha} (\varphi, \theta) = \Delta_{\alpha} \chi(\varphi, \theta),
\end{equation}
\begin{equation}
\Delta_{\alpha}  =   \sum_{\beta} \lambda_{\alpha \beta} \Delta_{\beta} \ln{\frac{2 \gamma \omega_D}{\pi T_c}\,}\,,
\label{gapeqlin}
\end{equation}
where
\begin{equation}
\lambda_{\alpha \beta} \equiv - V_{\alpha \beta} \nu_{\beta},
\label{intmatr}
\end{equation}
while the eigenvector $\Delta_{\beta}$ corresponding to the largest eigenvalue of the interaction matrix $\lambda_{\alpha \beta}$
determines the ratios between gaps on different Fermi surfaces 
set at $T_c$. Here $\nu_{\beta}$ are the densities of states (DOS) per one spin direction for various Fermi surfaces. While the
system of gap equations Eq.(\ref{gapeq}) seems to give a temperature-dependent ratio for gaps on
different Fermi surfaces, this result would violate the BCS logarithmic approximation.
Within the logarithmic accuracy, the ratios between the gaps on different Fermi surfaces is set at $T_c$ from Eq.(\ref{gapeqlin}),
and is a temperature-independent constant set by the interaction matrix Eq.(\ref{intmatr}),
the highest eigenvalue gap eigenvector \cite{geilikman,kresin}.
Then the multiband problem can be parameterized in terms of $T_c$ and the gap ratios, similar
to how this is done in the single-band BCS model. An additional related difficulty comes from the fact that
the kernel for the system of Fredholm integral equations Eq.(\ref{gapeq}) is asymmetric, i.e., 
$\lambda_{\alpha \beta} \neq \lambda_{\beta \alpha}$ for $\nu_{\alpha} \neq \nu_{\beta}$. 
Generalizing the approach of Pokrovskii\cite{pokrovskii} and following Geilikman \textit{et al.}\cite{geilikman}, 
we introduce new variables and a symmetric kernel 
$\mu_{\alpha \beta} = \mu_{\beta \alpha}$,
\begin{eqnarray}
\zeta_{\alpha} & = & \sqrt{\nu_{\alpha}} \Delta_{\alpha} \\ 
\mu_{\alpha \beta} & = & - \sqrt{\nu_{\alpha}} V_{\alpha \beta}  \sqrt{\nu_{\beta}}.
\end{eqnarray}
As usual, the gap equation in terms of the universal scale $T_c$ can be obtained by subtracting Eq.(\ref{gapeqlin}) from
Eq.(\ref{gapeq}).  Writing the result in new variables, we find:
\begin{widetext}
\begin{equation}
T \sum_{\beta n} \int d \xi \mu_{\alpha \beta} \zeta_{\beta} 
\left[\left\langle \frac{|\chi(\varphi, \theta)|^2}{\omega_n^2 + \xi^2 + |\Delta_{\beta}|^2 |\chi(\varphi, \theta)|^2}\,
\right\rangle_{\Omega} - \frac{1}{\omega_n^2 + \xi^2}\, \right] = 
\sum_{\beta} \mu_{\alpha \beta} \zeta_{\beta} \ln{\left[\frac{T}{T_c}\right]}.
\label{interstep}
\end{equation}
\end{widetext}
Let us now multiply Eq.(\ref{interstep}) by $\zeta_{\alpha}$, and sum over $\alpha$, using the symmetry of $\mu_{\alpha \beta}$
and Eq.(\ref{gapeqlin}),
\begin{equation}
\sum_{\alpha} \zeta_{\alpha} \mu_{\alpha \beta} =  \sum_{\alpha} \mu_{\beta \alpha} \zeta_{\alpha} = 
\frac{\zeta_{\beta}}{\ln{\frac{2 \gamma \omega_D}{\pi T_c}\,}\,}\,
\end{equation}
The gap equation is then considerably simplified, and can be written in the universal form:
\begin{widetext}
\begin{equation}
T \sum_{\beta n} \int d \xi u_{\beta}^2
\left[\left\langle \frac{|\chi(\varphi, \theta)|^2}{\omega_n^2 + \xi^2 + |\Delta_{\beta}|^2 |\chi(\varphi, \theta)|^2}\,
\right\rangle_{\Omega} - \frac{1}{\omega_n^2 + \xi^2}\, \right] =  \ln{\left[\frac{T}{T_c}\right]}.
\label{univgap}
\end{equation}
\end{widetext}
Here
\begin{equation}
u_{\beta}^2 = \frac{\zeta_{\beta}^2}{\sum_{\alpha} \zeta_{\alpha}^2}\, = 
\frac{\nu_{\beta} \Delta_{\beta}^2}{\sum_{\alpha} \nu_{\alpha} \Delta_{\alpha}^2}\,
\label{us}
\end{equation}
are constant ratios, determined by the gap eigenvector at $T_c$ and the corresponding densities of states.
The coefficients $u_{\beta}$ are automatically normalized:
\begin{equation}
\sum_{\beta} u_{\beta}^2 = 1.
\end{equation}
As demonstrated in the above derivation, all physical properties in the multiband BCS model can be expressed in terms 
of the transition temperature $T_c$, the relevant DOS, and temperature-independent gap amplitudes. 
The gap amplitudes themselves, however, are not universal. The ratios of the gaps on different Fermi 
surfaces are non-universal constants, determined
by the relevant interaction matrix (the highest-eigenvalue eigenvector for $\lambda_{\alpha \beta}$ in Eq.(\ref{gapeqlin})).  
Eq.(\ref{univgap}) can be easily solved at $T=0$. Introducing the usual normalization for the gap,
\begin{equation}
\Delta_{BCS} =  \frac{\pi T_c}{\gamma}\, e^{- \langle |\chi(\varphi, \theta)|^2 \ln{|\chi(\varphi, \theta)|} \rangle},
\label{BCSD}
\end{equation}
and defining $t_{\alpha}$ as
\begin{equation}
\Delta_{\alpha 0} \equiv t_{\alpha} \Delta_{BCS},
\label{gapt0}
\end{equation}
we find:
\begin{equation}
\sum_{\alpha} u_{\alpha}^2 \ln{t_{\alpha}} = 0,
\end{equation}
which sets a constraint on the amplitude of the gap eigenvector at $T=0$.  Eq.(\ref{BCSD}) gives
a standard expression for the zero-temperature amplitude of the energy gap for any single-band BCS superconductor.
For example,
\begin{eqnarray}
\Delta_{BCS} &=& \frac{\pi T_c}{\gamma}\, \simeq 1.76 T_c, \ \ \ s-wave, \\
\Delta_{BCS} &=& \frac{\sqrt{2} \pi T_c}{\gamma \sqrt{e}}\, \simeq 1.51 T_c, \ \ \ d-wave.
\label{swdw0tgap}
\end{eqnarray}
Note that due to the normalization condition Eq.(\ref{norm}), our definition for the d-wave gap amplitude is
different from the standard definition by a factor of $\sqrt{2}$.
The temperature dependence of the gap amplitude $\Delta(T)$ is determined by the universal gap equation Eq.(\ref{univgap}).
For the general case multiband case it differs from the single-band BCS temperature dependence:
\begin{widetext}
\begin{equation}
\ln{\frac{T}{T_c}\,} = \sum_{\alpha} \sum_{n=0}^{\infty} u_{\alpha}^2 
\left[
\left\langle\frac{|\chi(\varphi, \theta)|^2}{\sqrt{(n + 0.5)^2 + 
(t_{\alpha} \Delta(T)/2 \pi T)^2 |\chi(\varphi, \theta)|^2}}\,\right\rangle_{\Omega} - \frac{1}{n + 0.5}\, \right]
\label{gapTdep}
\end{equation}
\end{widetext}
Since the ratios of the gaps on different Fermi surfaces are set at $T_c$, one can obtain thermodynamic potential integrating
the gap equation Eq.(\ref{gapTdep}) over the single coupling constant, $T_c$:
\begin{equation} 
\Omega_S - \Omega_N = - \int_0^{T_c} \frac{d T_c'}{T_c'}\, \sum_{\alpha} \nu_{\alpha} \Delta_{\alpha}^2\left(\frac{T}{T_c'}\, \right).
\label{omegaa}
\end{equation}
Thermodynamics is then given by a sum of standard weak coupling expressions for
separate bands expressed in terms of temperature-dependent energy gaps $\Delta_{\alpha}(T)$:
\begin{widetext}
\begin{equation}
\Omega_S - \Omega_N = - \pi T \sum_{\alpha, n} \nu_{\alpha} \left\langle \sqrt{\omega_n^2 + \Delta_{\alpha}^2 |\chi(\varphi, \theta)|^2} + 
\frac{\omega_n^2}{\sqrt{\omega_n^2 + \Delta_{\alpha}^2  |\chi(\varphi, \theta)|^2}}\, - 2 |\omega_n| \right\rangle_{\Omega}
\label{omegab}
\end{equation}
\end{widetext}
Integrating the above expression over $\omega$ at zero temperature, one finds the familiar\cite{geilikman} factorized result
for the ground state energy:
\begin{widetext}
\begin{equation}
E_S - E_{N} = - \sum_{\alpha} \frac{\nu_{\alpha} \Delta_{\alpha 0}^2}{2}\, \langle |\chi(\varphi, \theta)|^2 \rangle_{\Omega} = 
-  \Delta_{BCS}^2 \sum_{\alpha} \frac{\nu_{\alpha} t_{\alpha}^2}{2}\,,
\label{energ} 
\end{equation}
\end{widetext}
where we have used the normalization condition for $\chi(\varphi, \theta)$ given by Eq.(\ref{norm}).
The multiband gap equation does produce overall change for the relevant quantities, such as, for example,
specific heat jump at transition temperature, $T_c$, obtained, for example, in Ref.\cite{geilikman}:
\begin{widetext}
\begin{equation}
\frac{\Delta C}{C}\, = \frac{12}{7 \zeta(3) \sum_{\alpha} \nu_{\alpha}}\, 
\frac{\sum_{\alpha} \nu_{\alpha} \Delta_{\alpha}^4}{\left(\sum_{\alpha} u_{\alpha}^2 \Delta_{\alpha}^2 \right)^2}\,
= \frac{12}{7 \zeta(3) \sum_{\alpha} \nu_{\alpha}}\,    
\frac{\left(\sum_{\alpha} \nu_{\alpha} \Delta_{\alpha}^2 \right)^2}{\sum_{\alpha} \nu_{\alpha} \Delta_{\alpha}^4}\,
\end{equation}
\end{widetext}
The above result of Geilikman \textit{et al.}\cite{geilikman} is applicable to any anisotropic or unconventional superconductors
belonging to a one-dimensional representation of the point group\cite{VG,gorkov}, which includes d-wave
multiband superconductors. For superconductors belonging to a degenerate representation of the point
group, the corresponding generalized formula is given by
\begin{equation}
\frac{\Delta C}{C}\, 
=  \left[\frac{\Delta C}{C}\,\right]_{BCS} \frac{1}{\sum_{\alpha} \nu_{\alpha}}\,   
\frac{\left(\sum_{\alpha} \nu_{\alpha} \Delta_{\alpha}^2 \right)^2}{\sum_{\alpha} \nu_{\alpha} \Delta_{\alpha}^4}\,,
\end{equation}
where $[\Delta C/C]_{BCS}$ is the weak coupling value of the specific heat jump at $T_c$ for a given multiband representation, given,
for example, by Kuznetsova and the author in Ref.\cite{KB}. 

As a simple example, let us consider the two-band case.  The expression for the constant ratio of the energy gaps on
different Fermi surfaces is then very easily obtained from Eq.(\ref{gapeqlin})
in terms of the interaction constants Eq.(\ref{intmatr}) (also see, for example, Ref.\cite{zhitomirsky}):
\begin{widetext}
\begin{equation}
\frac{\Delta_2 (T)}{\Delta_1 (T)}\, = \frac{\Delta_2 (T_c)}{\Delta_1 (T_c)}\, = \frac{2 \lambda_{12}}{\lambda_{22} - \lambda_{11} + 
\sqrt{(\lambda_{11} - \lambda_{22})^2 + 4 \lambda_{12} \lambda_{21}}}\, \equiv s.
\label{gapratio}
\end{equation}
\end{widetext}
The other relevant parameter of the BCS model is $u_1^2$, defined by Eq.(\ref{us}):
\begin{equation}
u_1^2 = \frac{\nu_1}{\nu_1 + \nu_2 s^2} = 1 - u_2^2.
\label{us1}
\end{equation}
Since the gaps on all Fermi surfaces have the same temperature dependence, let us introduce a normalized gap 
amplitude $\Delta$ as given by Eq.(\ref{gapt0}),
\begin{equation}
\Delta_{\alpha}(T) = t_{\alpha} \Delta(T), \ \ \Delta(T=0) = \Delta_{BCS}.
\end{equation}
We then easily find the expression for zero-temperature gaps on the two Fermi surfaces:
\begin{equation}
t_1 = s^{- u_2^2}, \ \ \ t_2 = s^{u_1^2}.
\end{equation}
The temperature dependence of the gaps on the two Fermi surfaces must be the same. In addition to $T_c$, thermodynamics
is completely determined by two other parameters,  $s$ and $u_1$, which can be easily found from experiment.

We conclude that the universality of the weak-coupling BCS-like model is applicable to the multiband case.
However, some non-universal constants, which depend on the interactions, do enter the problem as temperature-independent
parameters. The total number of independent parameters for an $m$-band BCS model is then significantly reduced to $2m - 1$
measurable constants: $T_c$, $m-1$ independent constant DOS ratios $\nu_{\alpha}/\nu_{\beta}$, and $m-1$ independent constant
gap ratios $\Delta_{\alpha}/\Delta_{\beta}$.  A significant temperature dependence of the energy gap ratios on different
Fermi surfaces as, for example, observed in MgB$_2$, then corresponds to the failure of the weak coupling 
logarithmic approximation.  

\section{Partial gapless state in magnetic field in s-wave superconductors.}

In this section we consider the unbalanced pairing problem\cite{LO,FF} for a multiband s-wave superconductor 
placed in paramagnetic magnetic field. The author and Gor'kov\cite{BG2} have recently shown that 
a new stable gapless state appears in small magnetic fields. The emergence of this state for s-wave superconductors
is accompanied by zero-temperature first order phase transition. The s-wave case can be solved analytically at $T=0$.
In what follows, we present our results for the stability of the gapless state in s-wave multiband superconductors 
and the details of the $B-T$ phase diagram. 

\subsection{The nature of the gapless state}

We start by considering the general solution for the BCS gap equation for a multiband superconductor
placed in magnetic field. We neglect the diamagnetic effects ($H_{c2}$).
The presence of paramagnetic magnetic field results in an additional  Pauli term in the Hamiltonian 
for each seperate band:
\begin{equation}
H_{p} = - \frac{1}{2}\, \sum_{\bm{k} i j \alpha} g_{\alpha} a^{\dagger}_{\alpha i} (\bm{k}) (\bm{I} \cdot \bm{\sigma}_{ij}) a_{\alpha j} (\bm{k}), 
\ \ \ \bm{I} \equiv \mu_B \bm{B}.
\label{pauli}
\end{equation} 
Here $g_{\alpha}$ is the g-factor for each band $\alpha$, taken to be isotropic in this section. 
The multiband gap equations in magnetic field have the same form Eq.(\ref{gapeq}), with the Gor'kov function $F$ given
by
\begin{widetext}
\begin{equation}
\hat{F}_{\alpha}(\omega_n, \bm{p})  = 
\frac{- i \hat{\sigma}^y \Delta_k(\bm{p})}{(i \omega_n - I \hat{\sigma}^z)^2 - \xi_{\alpha}(\bm{k})^2 - |\Delta_{\alpha}(\bm{p})|^2}\,
\end{equation}
\end{widetext}
The diagonalization procedure of Pokrovskii\cite{pokrovskii} and Geilikman \textit{et al.}\cite{geilikman} 
is applicable in the presence of arbitrary magnetic field, i.e., the ratios of the gaps on different Fermi
surfaces do not change as functions of  both temperature and magnetic field, and are determined by the gap eigenvector at $T_c$. 
For example, Eq.(\ref{gapratio}) for the two-band problem in the presence of magnetic field\cite{BG2} can be written
as:
\begin{widetext}
\begin{equation}
s \equiv \frac{\Delta_2 (T,B)}{\Delta_1 (T,B)}\, = \frac{\Delta_2 (T_{c0})}{\Delta_1 (T_{c0})}\, = 
\frac{2 \lambda_{12}}{\lambda_{22} - \lambda_{11} + 
\sqrt{(\lambda_{11} - \lambda_{22})^2 + 4 \lambda_{12} \lambda_{21}}}\,,
\label{gapratio1}
\end{equation}
\end{widetext}
where $T_{c0} = T_c(B=0)$. 

Thus, as it is the case with the energy gaps and thermodynamics, it is possible to obtain a complete solution
of the multiband problem in magnetic field in terms of $T_c$, gap ratios, and DOS ratios.
The linearized gap equations in magnetic field produce the instability curve $T_c(B)$,
which takes the following form:
\begin{widetext}
\begin{equation}
\ln{\frac{T_c}{T_{c0}}\,} = \Psi\left(\frac{1}{2}\,\right) - \sum_{\alpha} u_{\alpha}^2 Re\left[\Psi\left(\frac{1}{2}\, + i \frac{g_{\alpha} I}{4 \pi T_c}\, \right)\right],
\label{TcBcurve}
\end{equation}
\end{widetext}
where $u_{\alpha}$ are band-dependent constants given by Eq.(\ref{us}). Note that in the free electron case,
 $g_{\alpha} = 2$, the multiband instability curve Eq.(\ref{TcBcurve}) is reduced to the familiar single-band result,
in agreement with earlier results\cite{BG2}:
\begin{equation}
\ln{\frac{T_c}{T_{c0}}\,} = \Psi\left(\frac{1}{2}\,\right) - Re\left[\Psi\left(\frac{1}{2}\, + i \frac{I}{2 \pi T_c}\, \right)\right],
\label{TcBcurve1}
\end{equation}
The re-entrant behavior of $T_c$ with increased magnetic field normally indicates the possibility of first-order
phase transitions on the $B-T$ phase diagram. In a single-band problem the homogeneous gap equation gives rise to
a second unstable solution\cite{baltensperger} in magnetic fields close to the paramagnetic limit, also known as the 
Sarma\cite{sarma} state. The instability is resolved in favor of an inhomogeneous LOFF\cite{LO,FF} state. 
As we have shown in the
recent letter\cite{BG2}, in the two-band case one has three different solutions. In the general $m$-band case,
the number of solutions will be $m+1$. The $T=0$ solution of the gap equation will change form at $I = \Delta_{\alpha}(B)$.
For the s-wave case all solutions can be written out analytically as:
\begin{widetext}
\begin{equation}
\prod_{\alpha, g_{\alpha} I > 2 \Delta_{\alpha}(I)} 
\left(\frac{\sqrt{(g_{\alpha} I)^2 - 4 \Delta_{\alpha}(I)^2} + g_{\alpha} I}{2 \Delta_{\alpha 0}}\,\right)^{u_{\alpha}^2}
\prod_{\alpha, g_{\alpha} I < 2 \Delta_{\beta}(I)} \left(\frac{\Delta_{\beta}(I)}{\Delta_{\beta 0}}\, \right)^{u_{\beta}^2} = 1.
\label{magfieldsol}
\end{equation}
\end{widetext}
Here $\alpha$ and $\beta$ are band indices, $u_{\alpha}^2$ are given by Eq.(\ref{us}), while
$\Delta_{\alpha 0}$ correspond to the solution of the multiband gap equation at $T=0$ without the magnetic field,
Eq.(\ref{gapt0}). Stability of these solutions can be inferred from the ground-state energy, 
which can be easily obtained by the integration of the gap equation over the coupling constant $T_c$, as in Eq.(\ref{omegaa}):
\begin{widetext}
\begin{equation}
E_S - E_{N0} = - \sum_{\alpha} \frac{\nu_{\alpha} \Delta_{\alpha}(I)^2}{2}\, - 
\frac{1}{4}\, \sum_{\beta, g_{\beta} I > 2 \Delta_{\beta}(I)} \nu_{\beta} g_{\beta} I \sqrt{(g_{\beta} I)^2 - 4 \Delta_{\beta}(I)^2}. 
\label{energmag} 
\end{equation}
\end{widetext}
Here $E_{N0}$ is the normal-state energy in the absence of magnetic field. 

We can also write out the solution of the multiband gap equation at finite temperatures, since there is only a 
single temperature and field-dependent variable, the same for all $\Delta_{\alpha}(T,B)$: 
\begin{equation}
\delta(T,B) \equiv \frac{\Delta_{\alpha} (T, B)}{\Delta{\alpha}(T=0,B=0)}\,
\end{equation}
The gap equation then takes a simple form:
\begin{equation}
\ln{[\delta(T,B)]} = \sum_{\alpha} u_{\alpha}^2 f_0\left(\frac{\Delta_{\alpha} (T,B)}{T}\,, \frac{g_{\alpha} I}{2 T}\, \right),
\label{gapeqfinT}
\end{equation}
where
\begin{widetext}
\begin{equation}
f_0(x,y) = \int_0^{\infty} \frac{dt}{\sqrt{t^2+x^2}}\, \left(\frac{\sinh{(\sqrt{t^2+x^2})}}{\cosh{(y)} + \cosh{(\sqrt{t^2+x^2})}}\, - 1 \right),
\end{equation}
\end{widetext}
which is the same function that appears in the single-band model. The expression for thermodynamic
potential in magnetic field is an analytic continuation of the corresponding expression in zero field to 
$i \tilde{\omega_n} = i \omega_n - (g_{\alpha} I /2)$, which is also factorizable
\begin{widetext}
\begin{equation}
\Omega_S - \Omega_N(I) = - \frac{1}{2} \sum_{\alpha} \nu_{\alpha} |\Delta_{\alpha}(B,T)|^2 
\left(1 + f_1\left[\frac{\Delta_{\alpha}(B,T)}{T}\,,  \frac{g_{\alpha} I}{2 T}\,\right]\right),
\label{freeenT}
\end{equation}
with
\begin{equation}
f_1(x,y) = \frac{1}{2 x^2}\, \int_0^{\infty} t^2 dt \left(\cosh^{-2}{\left[\frac{\sqrt{t^2+x^2}-y}{2}\,\right]} +
 \cosh^{-2}{\left[\frac{\sqrt{t^2+x^2}+y}{2}\,\right]} -  \cosh^{-2}{\left[\frac{t-y}{2}\,\right]} - 
\cosh^{-2}{\left[\frac{t+y}{2}\,\right]}\right).
\end{equation}
\end{widetext}
Here
\begin{equation}
\Omega_N(I) = \Omega_{N0} - \frac{1}{4}\, \sum_{\alpha} \nu_{\alpha} (g_{\alpha} I)^2
\end{equation}
is the normal state energy in exchange magnetic field $I$.

\begin{figure}
\includegraphics[width=3.375in]{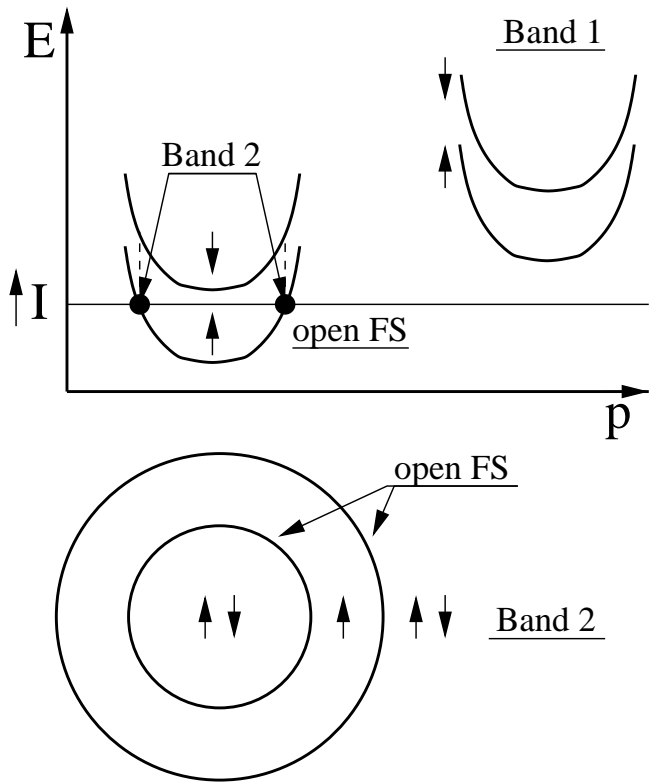}
\caption{Multiband partial Sarma state at $T=0$, characterized by fully polarized unpaired electrons near the Fermi surface
of the driven band. The state is stabilized by the presence of the gap on the primary Fermi surface.}
\label{sarmafig}
\end{figure}

From the general analysis of Eq.(\ref{energmag}) one can see that
the solution with $g_{\alpha} I > 2 \Delta_{\alpha}(I)$ for all $\alpha$ is always unstable, similar to the 
unstable Sarma state of the single-band problem\cite{baltensperger}. The low-field solution,
$g_{\alpha} I < 2 \Delta_{\alpha}(I) = 2 \Delta_{\alpha 0}$  for all $\alpha$, on the other hand, corresponds the multiband
BCS solution at $I=0$, which is stable in low enough magnetic fields. The other solutions of the gap equation
correspond to partial Sarma states, illustrated on Fig.\ref{sarmafig} for the two-band case and $g_{\alpha} = 2$. 
 The energy spectrum of the system for excitations near each FS is given by the poles of the 
Green's function $\hat{G}_{\alpha}(\omega_n, \bm{k})$:
\begin{equation}
 \hat{E}_{\alpha}(\bm{k}) = \sqrt{\xi_{\alpha}(\bm{k})^2 + |\Delta_{\alpha}(\bm{k})|^2} + \frac{1}{2}\, g_{\alpha} I \sigma^z 
\label{enspectrum}
\end{equation}
As the magnetic field exceeds the value of the smallest gap, a strip of unpaired fully polarized
quasiparticles forms in the vicinity of the corresponding Fermi surface, giving rise to a paramagnetic
magnetic moment. 
Similar to the single-band problem, the LOFF state competes with homogeneous solutions in high magnetic fields. 
A generalization of the instability curve Eq.(\ref{TcBcurve}) can be written for an arbitrary inhomogeneous 
q-vector as:
\begin{widetext}
\begin{equation}
\ln{\frac{T_c}{T_{c0}}\,} = \Psi\left(\frac{1}{2}\,\right) - \sum_{\alpha} u_{\alpha}^2 \left\langle Re\left[\Psi\left(\frac{1}{2}\, + i \frac{g_{\alpha} I + 2 (\bm{v}_{F \alpha} \bm{q}) }{4 \pi T_c}\, \right)\right] \right\rangle,
\label{TcBLOFF}
\end{equation}
\end{widetext}
where $\bm{v}_{F \alpha}$ is the Fermi velocity for band $\alpha$. $T_c(I)$ is then found as a maximum with respect to $\bm{q}$. 
The high-field phase transition is always second order. As the field is lowered, the regions
of stability of various phases and the exact nature of the LOFF state in the general case, when the energy gaps are of the 
same order of magnitude, can only be obtained numerically. 

The zero-temperature phase transition in partially gapless state in s-wave multiband superconductors is accompanied by the
appearance of the paramagnetic moment\cite{BG2}. In non s-wave multiband superconductors, however, magnetic moment
is always present in low magnetic fields due to the nodes in the energy spectrum. The transition to partial Sarma
state then corresponds to a crossover from the nodal regime to a regime with a full open Fermi surface. In completely
isotropic situation, the zero-temperature phase transition to partial Sarma state in s-wave superconductors is always 1-st order,
corresponding to the appearance of a finite paramagnetic magnetic moment.
Effects of spin-orbit, non-spherical Fermi surface, or gap anisotropy can turn this phase transition
into a smooth crossover. Similarly, due to effects of gap anisotropy, the first-order zero-temperature phase transition in 
multiband d-wave superconductors exists only in a certain region of parameters\cite{BG2}. The first-order phase transition,
if present, disappears above a certain critical temperature, $T_{cr}$.
 
\begin{figure}
\includegraphics[width=3.375in]{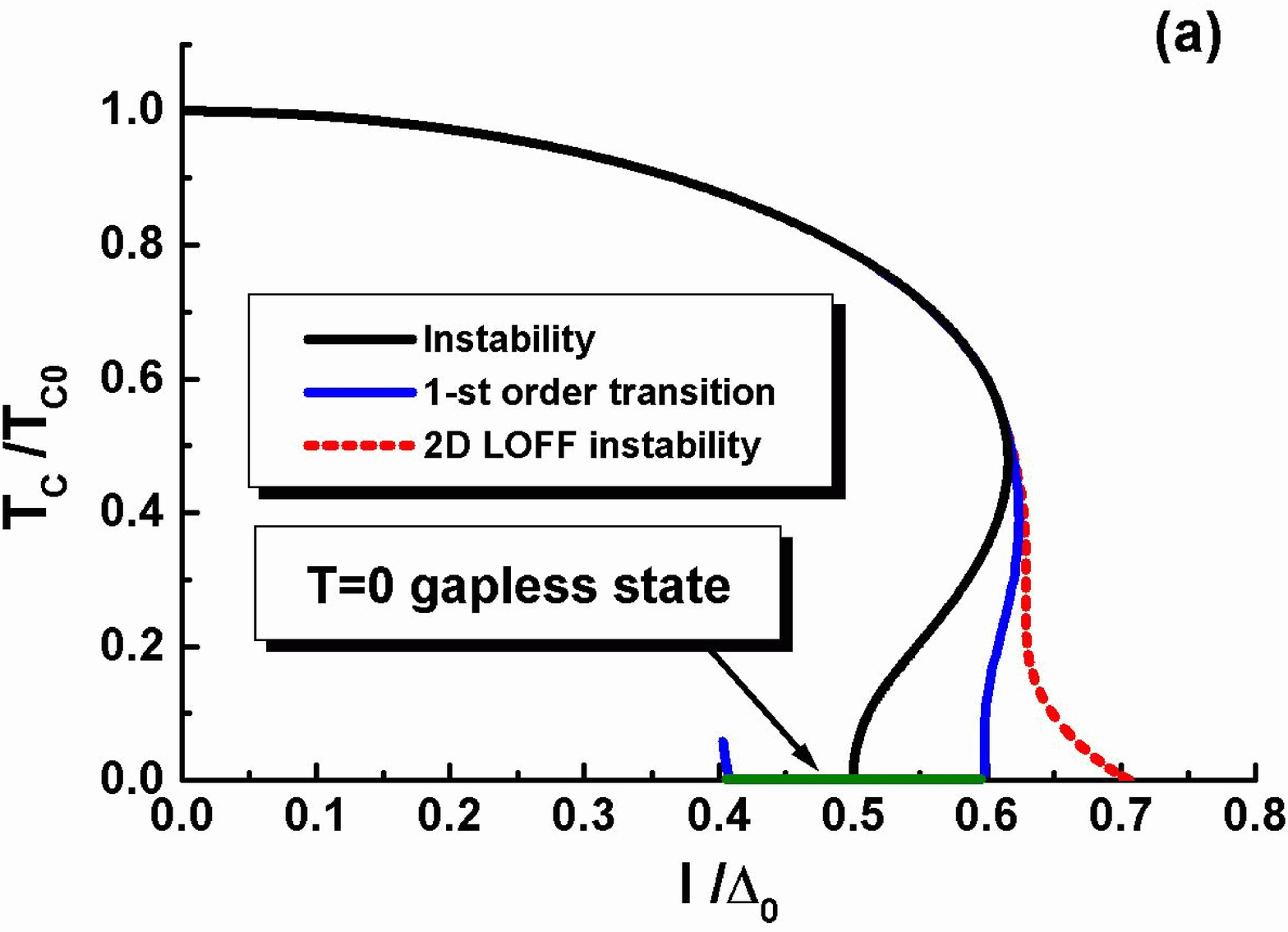}
\includegraphics[width=3.375in]{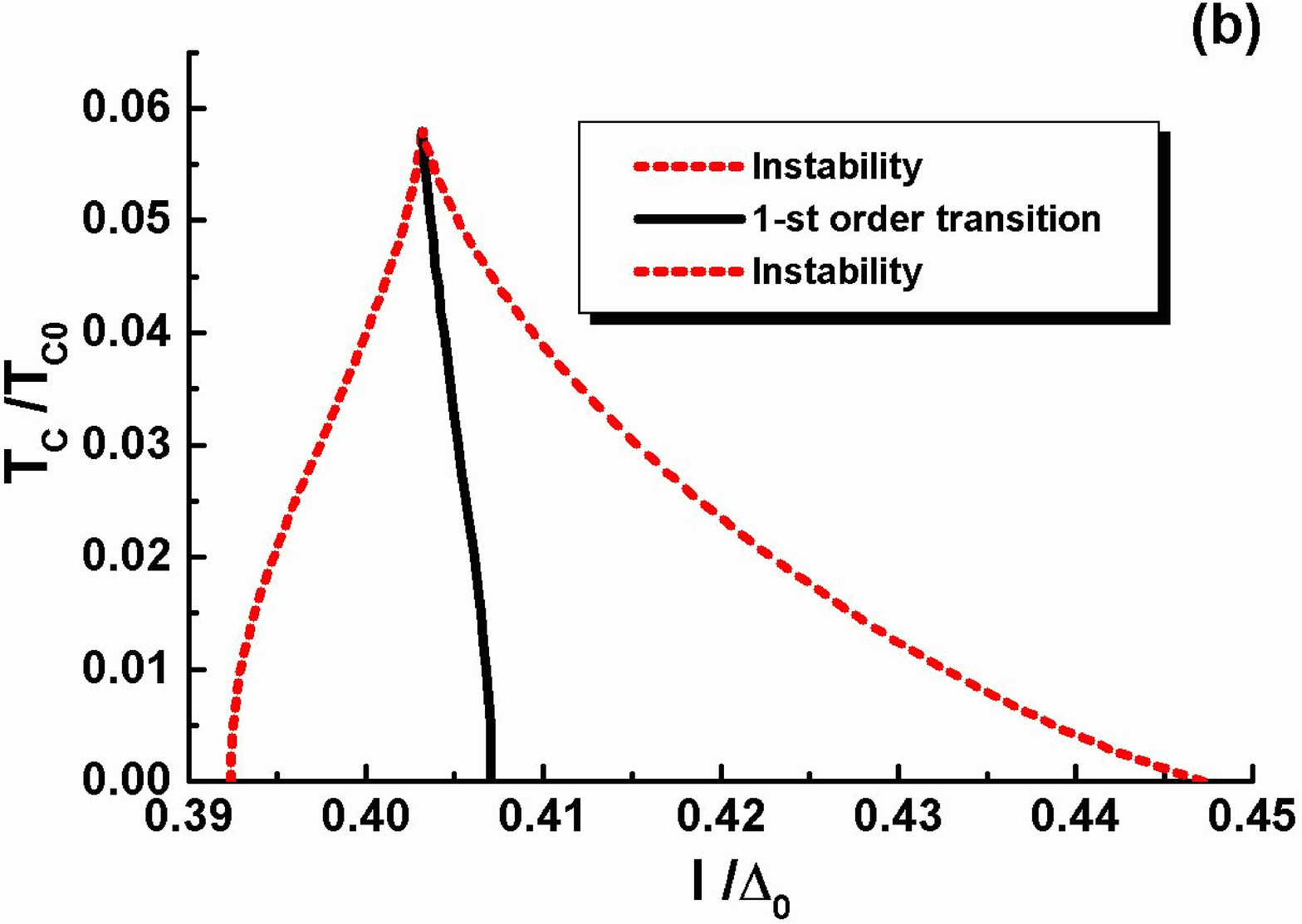}
\caption{(a) $B-T$ phase diagram for a 2D two-band superconductor with two equal circular Fermi surfaces 
($p_{F1} = p_{F2}$) and $(\Delta_2/\Delta_1) = 0.2$, $(m_2/m_1) = 25$. The blue line corresponds to the first order 
phase transition into the partial Sarma state from the normal state (right side, the Clogston limit) or fully gapped 
superconducting state (left side). The LOFF instability is shown by a dashed red line. The instability for the 
normal/uniform superconducting state is shown by the black line. (b) Details of low temperature first order transition from
fully gapped to gapless superconducting state. The black line marks the first order phase transition. The red dashed
lines correspond to the boundaries of the region of metastable states, where the hysteretic behavior is expected
to occur.}
\label{fig2}
\end{figure}

The analysis of the $B-T$ phase diagram for multiband superconductors depends on the same number of parameters
as the analysis of other properties considered in previous section. Thus, for example, energetic stability of various
gapless state will depend on the value of these additional parameters. We thus consider for
simplicity a two-band model, which has only two additional physical constants, the ratio of the densities of states
on two Fermi surfaces, $\nu_2/\nu_1$, and the ratio of the energy gaps $\Delta_2/\Delta_1$. We also assume
for simplicity $g_1 = g_2 = 2$ and only consider homogeneous solutions, since the LOFF state will depend
on the shape of both Fermi surfaces. Fig. \ref{fig2}a shows an example $B-T$ phase diagram for homogeneous
phases for the two-band isotropic s-wave model in exchange field and parameters $(\Delta_2/\Delta_1) = 0.2$ and $(\nu_2/\nu_1) = 25$
obtained with the help of the above expressions for the temperature-dependent gaps and thermodynamic potential
$\Omega_S$. The instability line in Fig. \ref{fig2}a for the transition into a uniform partially gapless superconducting
state has the characteristic reversal behavior, which indicates the presence of either a LOFF state\cite{LO,FF} or
a first order phase transition. The LOFF instability line for two equal circular Fermi surfaces with different masses in two
dimensions is shown by the red dashed line. When only the uniform superconducting states are considered, the
the partially gapless state will be separated from the normal state in high magnetic field and low temperatures 
by a first order Clogston-type transition. At low fields it is separated from the fully gapped superconducting state 
by a first order line that ends in a critical point. Fig. \ref{fig2b} shows the details of the first order phase
transition from fully gapped superconducting state to partially gapless superconducting state. The first order transition
is shown by the solid black line that ends in a critical point at $T=T_{CR} \simeq 0.058 T_{c0}$. The dashed red lines
mark the boundaries metastable region in the vicinity of the first order phase transition where unstable solutions 
of the gap equation are present.

\begin{figure}
\includegraphics[width=3.375in]{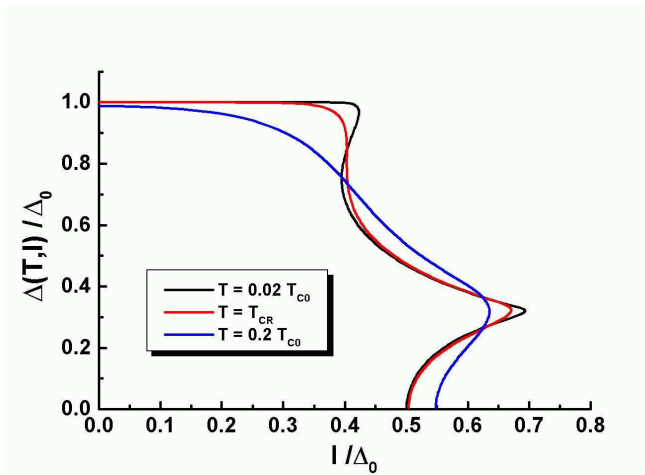}
\caption{Numerical solution of the two-band BCS gap equation $\Delta(T,B)$ at different temperatures for parameters in 
Fig.\ref{fig2}. Three solutions for the gap equation at low fields and temperatures reflect the existence of unstable region
near the first order phase transitions from fully gapped state into partially gapless state.
As the temperature is raised above the critical temperature, the first order region disappears, and there is only one
solution for the gap equation in low fields. The high-field behavior corresponds to the usual first order unstable
Clogston limit.}
\label{fig3}
\end{figure}

\begin{figure}
\includegraphics[width=3.375in]{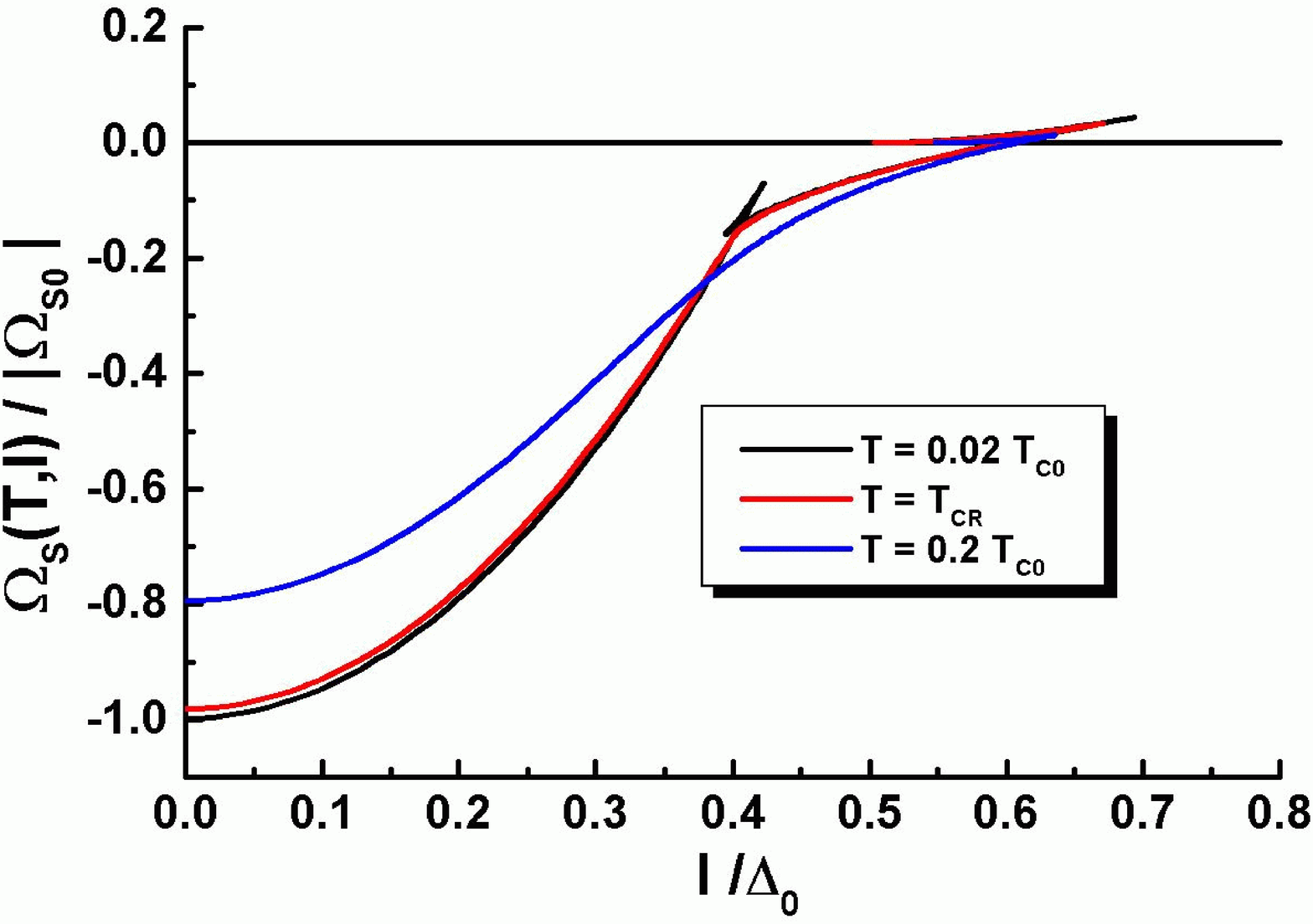}
\caption{Free energy $\Omega_S$ of the superconducting state for a two-band superconductor in magnetic field at different
temperatures for the parameters in Fig.\ref{fig2}. The reversal behavior at low temperatures $T < T_{cr} = 0.058 T_{c0}$ 
indicates the presence of the first order phase transition from the fully gapped superconducting state
into a partially gapless gapless superconducting state. The first order transition from the partially gapped superconducting
state to normal state in high magnetic fields happens at the point when the condensation energy $\Omega_S(T,I) = 0$.}
\label{fig4}
\end{figure}

Fig. \ref{fig3} shows the solution of the multiband gap equation Eq.(\ref{gapeqfinT}) as a function of temperature and magnetic
field. The unusual reversal behavior of the energy gap reflects the presence of a first order phase transition.
While there are three different solutions at a given field in a region near this phase transition, only one of these 
solutions is stable. Such a region corresponds to the region of the first order phase transition where the hysteresis exists. 
The stable solution corresponds to the minimum
of the Free energy, Eq.(\ref{freeenT}), shown in Fig. \ref{fig4}. The shape of the Gibbs Free energy at 
$T < T_{cr}=0.058 T_{c0}$ also reflects the presence of the metastable states near the first order phase transition into
the gapless superconducting state. The system always picks the lowest energy. Thus, the free energy of the system
as a function of magnetic field has a slope change at the point of first order phase transition, where the energy
gap and paramagnetic magnetization have a corresponding jump.


The zero-temperature phase transition into the new state is characterized by a metamagnetic jump of magnetization. At $T=0$
the magnetic moment appears sharply, from $M = 0$ to $M \neq 0$. At a finite temperature small magnetization is
present due to thermal population of the band with smaller energy gap. The first order phase transition line disappears
at $T>T_{cr}$, as the quasiparticle states above the gapless state become thermally populated. 
Thus, one has two metamagnetic transitions shown in Fig. \ref{fig3},
one corresponding to the transition into partially gapless Sarma state, the other is the first order phase 
transition from the partially gapped Sarma state into the normal state.

\begin{figure}
\includegraphics[width=3.375in]{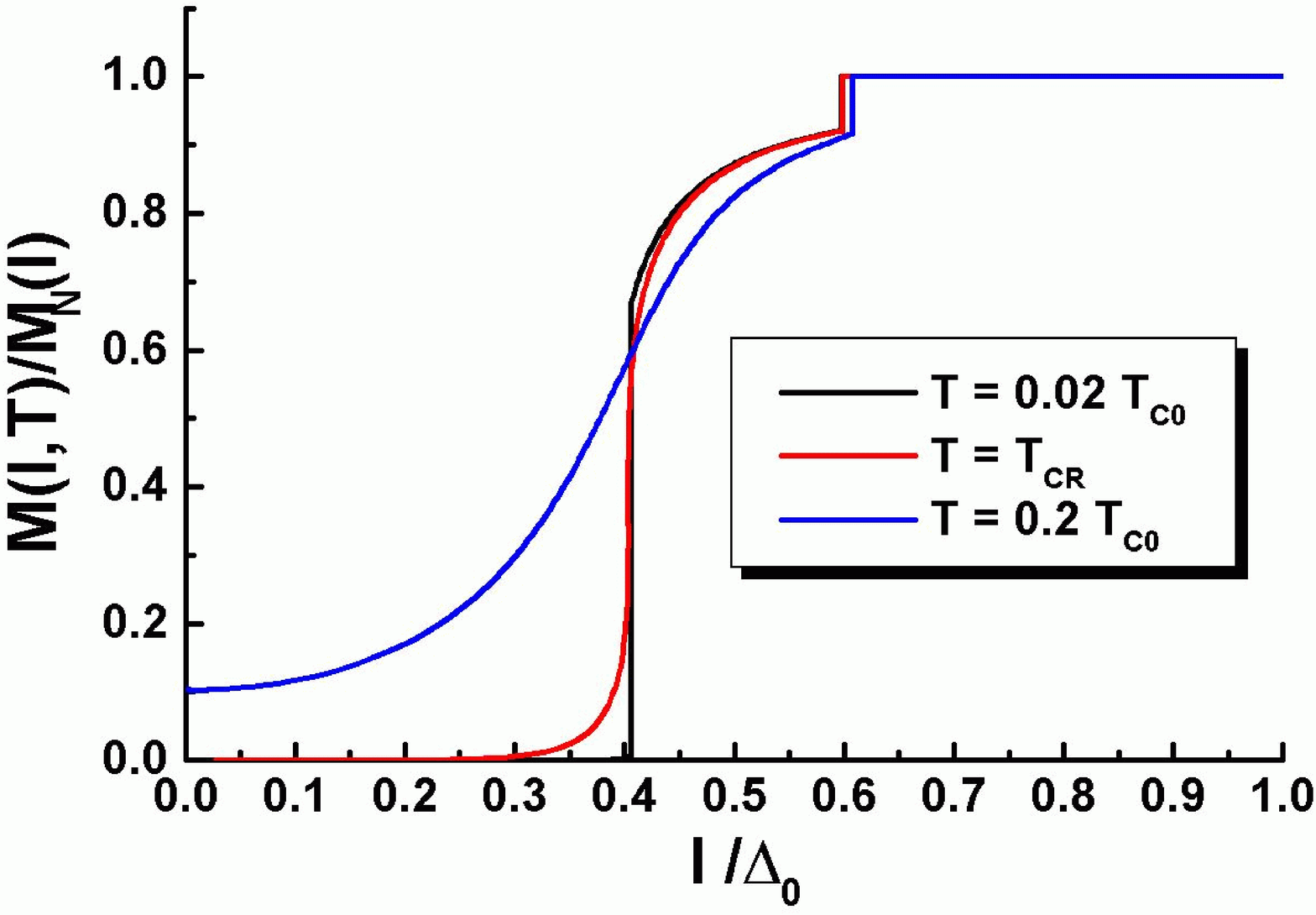}
\caption{Magnetic moment normalized to magnetic moment in the normal state at different temperatures for the parameters
in Fig. \ref{fig2}. The first order jump in magnetization from zero to finite magnetic moment in the
lower fields and temperatures $T < T_{cr} = 0.058 T_{c0}$ corresponds to a transition into partially gapless state. 
The sharp transition is absent at temperatures above the critical temperature, as quasiparticle states above the second 
smaller gap become thermally populated.}
\label{fig5}
\end{figure}

The analysis of the above equations can be done numerically for any parameters $\nu_2/\nu_1$ and 
$\Delta_2/\Delta_1$. For the s-wave case, however, analytical results for the energy and the magnetic moment
can be obtained at $T=0$. In a special case when the critical point is located at low enough temperatures,
the critical region is fully determined by the second Fermi surface and depends only on one energy scale.
We consider these simplified results below.

Finally, we note that the partial gapless state is not always present on the $B-T$ phase diagram; for 
some parameters of the two-band model, as shown below, the first order transition happens directly from the uniform
superconducting state into the normal state. 
      
\subsection{Magnetic properties of the partial gapless state at $T=0$}

In this section we consider the low field transition to partial gapless state. 
As we have seen in the previous section, the gap equation for the two-band model has three different
solutions and depends on two constant parameters, the ratio of the energy gaps on the two Fermi
surfaces $s$, determined only by interactions, Eq.(\ref{gapratio1}), and the ratio of the densities of 
states for the two bands. However, analytical expressions are more conveniently written in terms of
$s$ and a combination of these parameters $u_2^2$, defined in Eq. (\ref{us1}). 

The uniform low-field solution at $T=0$ is the same as the solution without the magnetic field.
The first order phase transition appears near $I \simeq \Delta_{20}$. It is thus convenient
to rewrite the solution in terms of $\Delta_{20}$ and other parameters of the second Fermi surface. For example,
the energy of the superconducting state at $T = 0$ and $B = 0$ is

\begin{equation}
E_{s0} = - \frac{1}{2 u_2^2}\, \nu_2 \Delta_{20}^2.
\end{equation}

Introducing

\begin{eqnarray}
\tilde{\Delta} & \equiv & \frac{\Delta_2(B,T)}{\Delta_{20}}\, = \frac{\Delta_1(B,T)}{\Delta_{10}}\, , \\
\tilde{I} & \equiv & \frac{I}{\Delta_{20}}\,
\end{eqnarray}

we find a simple expression for $I$ in the partially gapped state:

\begin{equation}
\tilde{I} = \frac{1}{2}\, \tilde{\Delta} \left(\tilde{\Delta}^{u_2^{-2}} + \tilde{\Delta}^{- u_2^{-2}}\right),
\end{equation}
or
\begin{equation}
\tilde{I} = \tilde{\Delta}  \cosh{(u_2^{-2} \ln{\tilde{\Delta}})}.
\label{gappp}
\end{equation}

The expression for the energy also simplifies:

\begin{equation}
\frac{E_S}{E_{S0}}\, = \tilde{\Delta} + 2 u_2^2 \tilde{I} \sqrt{\tilde{I}^2 - \tilde{\Delta}^2}
\end{equation}
The 1-st order transition point is determined from the smallest $\tilde{\Delta} < 1$ solution of the 
transcendental equation,
\begin{equation}
1 - u_2^2 \sinh{(2 u_2^{-2} \ln{\tilde{\Delta}_{cr}})} = \tilde{\Delta}_{cr}^{-2},
\label{tranpoint}
\end{equation}
where the solution changes abruptly from $\tilde{\Delta} = 1$ to the smaller gap branch of Eq.(\ref{gappp}).
The magnetization in the partial Sarma state as a function of $\tilde{\Delta}$ is also easily determined,
since it is also just a fraction of the normal state magnetization on the second Fermi surface:
\begin{equation}
M = - 2 \mu_B \nu_2 I \tanh({u_2^{-2} \ln{\tilde{\Delta}}}).
\label{mommm}
\end{equation}
Eqs. (\ref{gappp}) and (\ref{mommm}) determine the field dependence of magnetization in the partial gapless
state parametrically. Magnetization has a jump from $0$ to a finite value at the first order 
phase transition $\tilde{\Delta}_{cr}$. The transition point can be easily found analytically in a particular
case of $u_2^2 \ll 1$, corresponding to "induced" superconductivity on the second Fermi surface\cite{BG2}.
Indeed, introducing
\begin{equation}
\tau_{\Delta} \equiv \tilde{\Delta} - 1,
\end{equation}
Eq.(\ref{gappp}) transforms to:
\begin{equation}
\tau_I = \frac{1}{2}\, \frac{\tau_{\Delta}^2}{u_2^4}\, + \tau_{\Delta},
\label{gapT0}
\end{equation}
while for the condensation energy we easily find the following expression:
\begin{equation}
\frac{E_S}{E_{S0}}\, = 1 - 2 \tau_{\Delta}^2 - \frac{4}{3 u_2^4} \tau_{\Delta}^3,
\label{enT0}
\end{equation}
The cubic terms in the energy lead to a first order phase transition\cite{BG2} at 
\begin{equation}
\tau_{Icr} = - 3 u_2^4/8,
\end{equation}
where the  energy gap $\tau_{\Delta}$ changes abruptly from $\tau_{\Delta} = 0$ to $\tau_{\Delta} = - 3 u_2^4/2$.
The magnetization changes abruptly at $\tau_{Icr}$ from zero to
\begin{equation}
M_{cr} = \frac{3}{2}\, u_2^2 \mu_B \nu_2 \Delta_{20}.
\end{equation} 

\subsection{High-field Clogston limit and energetic stability of the gapless state at $T=0$.}

We now turn to the first order transition in high magnetic field, or the modification for the 
Clogston criterion for the two-band model. The standard Clogston Criterion involves a transition
from the uniform superconducting state to the normal state, determined by
\begin{widetext}
\begin{equation}
E_{S0} = - \frac{\nu_1 \Delta_{10}^2}{2}\, - \frac{\nu_2 \Delta_{20}^2}{2}\, = E_{N}(I) = - (\nu_1 + \nu_2) I_{clog}^2
\end{equation}
\end{widetext}
We thus obtain
\begin{equation}
I_{clog} = \frac{1}{\sqrt{2}}\, \sqrt{\frac{\nu_1 \Delta_{10}^2 + \nu_2 \Delta_{20}^2}{\nu_1 + \nu_2}\,}
\end{equation}
The transition first order transition from gapped two-band superconductor to the normal state
only happens at $I = I_{clog}$ in the absence of the partially gapless state. Since the 
partially gapless state is different from the uniform gapped state, the high field transition
to the normal state will also happen in a different magnetic field, now determined by
\begin{equation}
E_S(I) =  E_N(I)
\end{equation}
After simple calculations, we find:
\begin{equation}
\tilde{\Delta}^{2u_2^{-2}}_{clog1} = \frac{u_1 s^2}{u_1 (1-s^2) + \sqrt{u_1^2 - 2 s^2}}\,,
\label{clg}
\end{equation}
where $u_1 = \sqrt{1 - u_2^2}$, and the new Clogston limit $I_{clog1}$ then determined by Eq.(\ref{gappp})
that connects $\tilde{\Delta}$ and $I$ in the partially gapless state. 

\begin{figure}
\includegraphics[width=3.375in]{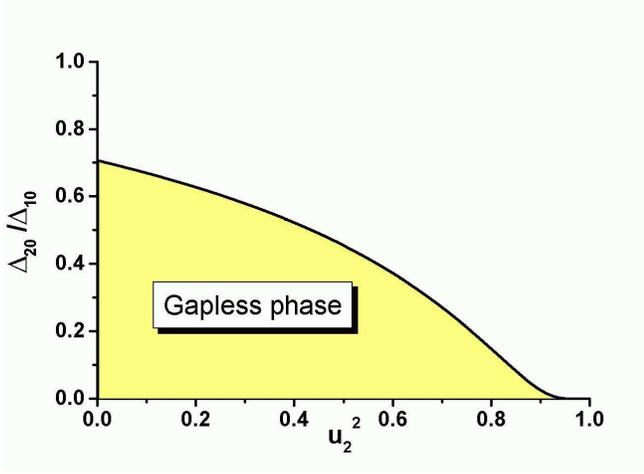}
\caption{The region of parameters for the two-band model for which the $T=0$ gapless phase is present.}
\label{fig6}
\end{figure}

It is obvious that the condition
\begin{equation}
I_{clog1} > I_{cr},
\label{condit}
\end{equation}
where $I_{cr}$ is the magnetic field for the $T=0$ transition from uniform into partially gapless
state, must be met for the partially gapless state to be present on the phase diagram. In particular,
it is obvious that for
\begin{equation}
s > \frac{u_1}{\sqrt{2}}  
\end{equation}
the partially gapless state is definitely not present. The accurate condition is given by Eq.(\ref{condit}), which
upon substitution of Eq.(\ref{clg}) into Eq.(\ref{tranpoint}) takes the following form:
\begin{equation}
1 - u_2^2 \sinh{(2 u_2^{-2} \ln{\tilde{\Delta}_{cr}})} > \tilde{\Delta}_{cr}^{-2}.
\label{condit1}
\end{equation}
The region of parameters of the two-band model where the partially gapless state exists given by Eq.(\ref{condit1})
is shown in Fig. \ref{fig6}. We see that in order for the gapless state to be present and easily observable, the second
gap has to be much smaller than the first gap, while the density of states on the second Fermi surface should be quite
a bit larger than on the first one, otherwise the entropy change associated with the first order transition will be
very small and not easily detectable. Such transition will also be turned into a crossover in the presence of even
small gap anisotropy, impurities, non-zero temperature, or other effects\cite{BG3}.

We have not considered the details of the high-field stripe LOFF\cite{LO,FF} region beyond the LOFF instability in
the normal state given by Eq.(\ref{TcBLOFF}) and shown in Fig. \ref{fig2}, since this calculation 
depends on the details of the shape of the Fermi surfaces, their dimensionality, and thus a number of additional 
parameters\cite{BR}. Unlike the Clogston limit, the LOFF region is very non-universal and 
sensitive to defects. We note, however, that the LOFF stripe phase will also be unusual, since it
forms in high enough magnetic fields in the vicinity of the first order transition between partially gapless state 
and  the normal state. In particular, the inhomogeneous LOFF state will likely involve superconducting stripe order
made out of the partially gapless state. It is also not completely clear whether the phase transitions between 
the partially gapless superconducting state and the LOFF state and between the LOFF state and the normal state
will be first or second order. 

\subsection{The low-field critical region.}

The mathematics of the low-field critical region near the $T=0$ first order phase transition 
into the partially gapless superconducting state and the associated thermodynamics is rather bulky, 
and can only be studied numerically for the general case (see Figs.\ref{fig2}-\ref{fig5}). 
Nevertheless, as it was shown in the previous subsections,
\textit{only parameters for the second Fermi surface} are relevant for the first order
phase transition at $T=0$. Unfortunately, \textit{both} Fermi surfaces
determine the critical region at finite temperatures. However, when $u_2^2 \ll 1$, the critical region lies at very low
temperatures, and thus only the quasiparticle excitations near the gapless or nearly gapless state on the 
second Fermi surface are important. The results for the critical region then depend only on 
parameters of the second Fermi surface only, and thus it can be found in a universal form. 
We note that the special
case $u_2^2 \ll 1$, corresponding to superconductivity driven by a single Fermi surface and induced by
interactions on other Fermi surfaces is, perhaps, most common for multiband superconductors.
In this section we consider such a weak first order transition, assuming that both Fermi surfaces are completely isotropic.

The $T=0$ first order phase transition in this limit happens near $I \simeq \Delta_{20}$. To describe this phase 
transition and thermodynamics near it, following Ref. \cite{BG2}, it is convenient to introduce new dimensionless 
variables that correspond to deviation of the second energy gap and the magnetic field from $\Delta_{20}$:

\begin{equation}
\tau_I \equiv \frac{I - \Delta_{20}}{\Delta_{20}}\,,
\label{tauIdefin}
\end{equation}

\begin{equation}
\tau_{\Delta} \equiv \frac{\Delta_2(T,I) - \Delta_{20}}{\Delta_{20}}\,,
\label{tauDdefin}
\end{equation}

and dimensionless temperature

\begin{equation}
t \equiv \frac{T}{\Delta_{20}}\,.
\end{equation}

The solution for the metamagnetic transition at $T=0$ in this limit is given by Eqs.(\ref{gapT0}),(\ref{enT0}).
It is not difficult to extend this solution to finite temperatures. The gap equation then takes 
the following form:

\begin{equation}
\tau_{\Delta} = - u_2^2 \sqrt{t}  \int_0^{\infty} 
\frac{\sqrt{x} dx}{\cosh^2{\left[x + \frac{\tau_{\Delta} - \tau_I}{2t}\,\right]}}\,.
\label{gapT}
\end{equation}

Thermodynamics near the critical point is simply given by the integral of the gap equation,
\begin{equation}
\Omega_S - \Omega_{S0} = 4 |\Omega_{S0}| \int_{- \infty}^{\tau_I} \tau_{\Delta}(\tau_I) d \tau_I.
\end{equation} 
The critical point can be found by differentiating Eq.(\ref{gapT}) and solving the following two equations:
\begin{equation}
\frac{d \tau_I}{d \tau_{\Delta}}\, = \frac{d^2 \tau_I}{d \tau_{\Delta}^2}\, = 0.
\end{equation} 

The critical temperature is given by:
\begin{equation}
t_{cr} = A_1^2 u_2^4,
\end{equation}
with
\begin{equation}
A_1 \equiv \int_0^{\infty} \frac{\sqrt{x} \tanh{(x+A_2)} dx}{\cosh^2{(x+A_2)}}\,,
\end{equation}
where $A_2$ is the solution of 
\begin{equation}
\int_0^{\infty} \sqrt{x} dx \frac{1 -2 \sinh^2{(x+A_2)}}{\cosh^4{(x+A_2)}}\, = 0.
\end{equation}
The energy gap and magnetic field at the critical point are given by
\begin{eqnarray}
\tau_{\Delta,cr}  & = &  - A_1 u_2^4 \int_0^{\infty} \frac{\sqrt{x} dx}{\cosh^2{(x+A_2)}}\, \\
\tau_{I,cr} & = &  \tau_{\Delta,cr} - A_2 t_{cr}.
\end{eqnarray}
Solving for $A_1$ and $A_2$ numerically, we obtain:
\begin{eqnarray}
t_{cr} & = & 0.3129 u_2^4, \\
\tau_{\Delta,cr} & = & - 0.7541 u_2^4, \\
\tau_{I,cr} & = & - 0.4071 u_2^4.
\end{eqnarray}

It is not difficult to see that the solutions for the gap equation, the energy, and the magnetic moment 
in the critical region depend on single energy scale $T_{cr}$, or $u_2^4 \Delta_{20}$. Thus, universal
numerical results for the critical region can be obtained. 

\begin{figure}
\includegraphics[width=3.375in]{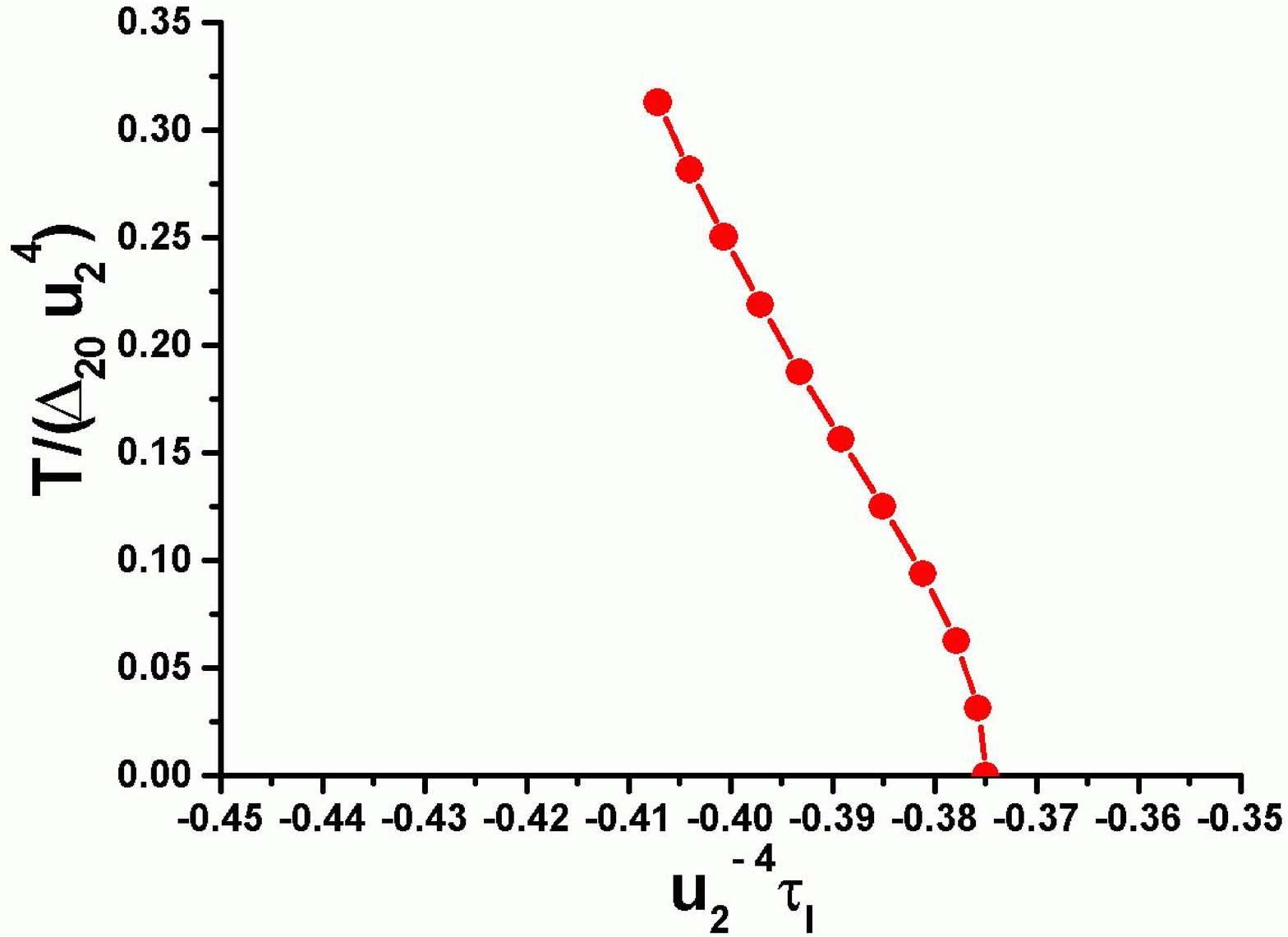}
\caption{The $T-B$ phase diagram near the first order transition to partially gapless Sarma state
in the special case of weak first order transition, marked by a red solid line.}
\label{fig7}
\end{figure}

\begin{figure}
\includegraphics[width=3.375in]{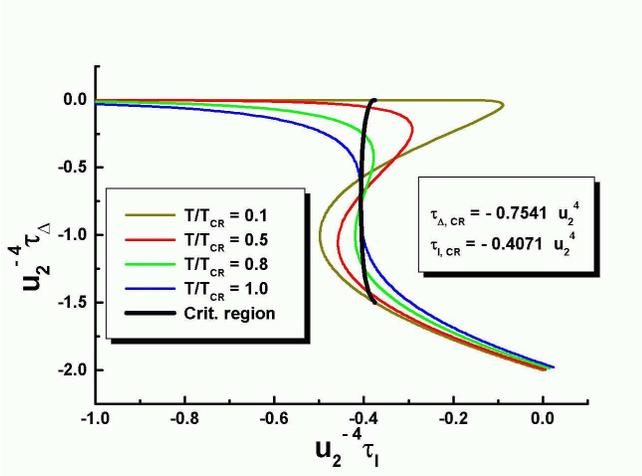}
\caption{Solution for the gap equation in the low field critical region as a function of magnetic field and temperature.
The black line is the first order transition line for different temperatures in $B-\Delta$ coordinates, also shown in 
Fig.\ref{fig7} in $B-T$ coordinates.}
\label{fig8}
\end{figure}

\begin{figure}
\includegraphics[width=3.375in]{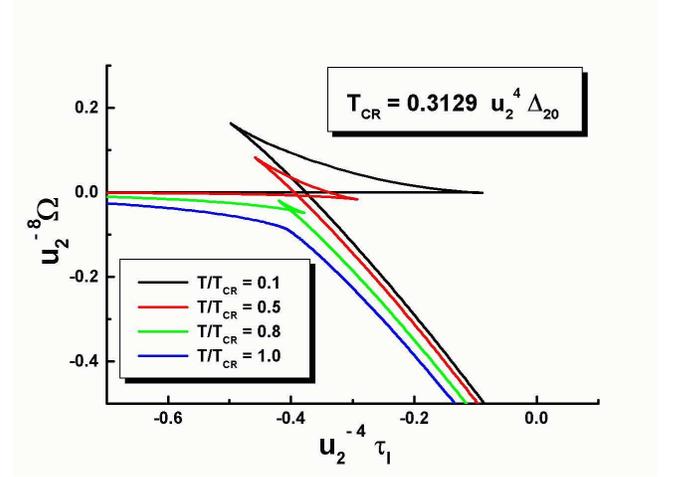}
\caption{Gibbs free energy of the superconducting state near the first order phase transition to the partially gapless state
for different temperatures $T \le T_{cr}$. The unusual behavior of the Gibbs potential indicates the presence of metastable
states near the first order transition from fully gapped to partially gapless superconducting state.}
\label{fig9}
\end{figure}

\begin{figure}
\includegraphics[width=3.375in]{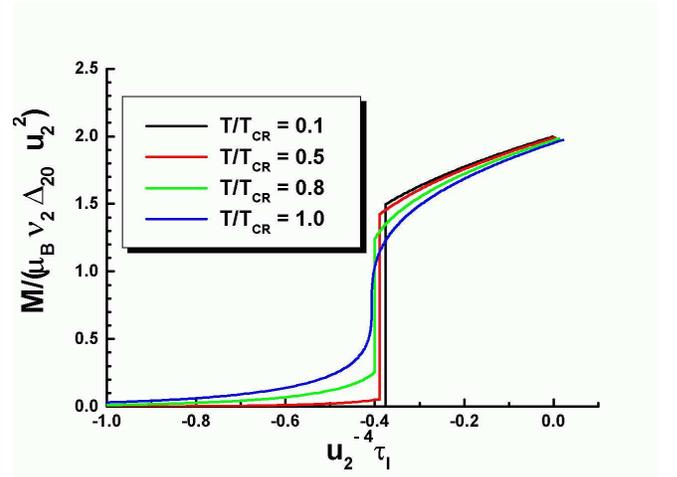}
\caption{First order metamagnetic transition into partially gapless state at different temperatures $T < T_{cr}$. The
transition disappears and becomes a crossover at $T > T_{cr}$.}
\label{fig10}
\end{figure}

In Fig. \ref{fig7} we show the first order transition line on the $B-T$ phase diagram in universal 
units $u_2^{-4} \tau_I$ and $T/(\Delta_{20} u_2^4)$. Fig. \ref{fig8} shows the solution of the 
gap equation in the critical region at different temperatures in universal units  $u_2^{-4} \tau_I$ 
and $u_2^{-4} \tau_{\Delta}$. The gap equation has three different solutions. The black line is
the line of first order phase transitions of Fig. \ref{fig7} in these coordinates. The two 
$\tau_{\Delta}$ on this line at a given field $\tau_I$ correspond to the two coexisting superconducting states on 
the first order line at a given temperature. The thermodynamic potential $\Omega = (\Omega_S/\Omega_{S0}) - 1$
is shown in reduced units $\Omega u_2^{-8}$ in Fig. \ref{fig9}. The behavior of the gap as a function of magnetic
field results, as usual, in the presence of metastable states in the vicinity of the first order phase transition.
Finally, Fig.\ref{fig10} shows the jump in paramagnetic magnetization at different temperatures below 
the critical temperature $T_{CR}$ in reduced units of $M/(\mu_B \nu_2 \Delta_{20} u_2^2)$. As the temperature
is raised to $T_{CR}$, the first order jump disappears.

\section{Conclusions.}

Let us now briefly summarize our main results. We have performed a detailed calculation
of energetically stable homogeneous superconducting states in the unbalanced pairing problem 
for s-wave multiband superconductors. Our analysis shows that this problem differs from one for 
single-band superconductors, and
that qualitatively new partially gapless states may be present in the low temperature
region of the $B-T$ phase diagram. The new states are characterized by a gapless Fermi spectrum,
open Fermi surfaces, and a finite paramagnetic magnetic moment. The phase transition between
fully gapped and gapless superconducting states in magnetic field at $T=0$ is a metamagnetic first order phase transition,
which corresponds to a sharp jump in magnetization on one of the Fermi surfaces that becomes gapless.

The superconducting order is present on both Fermi surfaces, as shown in Fig. \ref{sarmafig}.   
At finite low temperatures the metamagnetic first transition results a first order line on the $B-T$ phase
diagram that ends in a critical point, as shown in Fig.\ref{fig2}. The presence of the gapless superconductivity
also modifies the high-field Clogston limit. While the new state is analogous to the one 
studied in Ref.\cite{LW}, it is not the same. Unlike the situation encountered in Bose condensation or
high energy physics, unbalanced pairing of different species of fermions in superconductors is energetically
unfavorable because of the large difference of the corresponding Fermi surfaces. Pairing in multiband superconductors
is associated with each Fermi surface separately, although the gaps on different
Fermi surfaces are related by the interaction. In the weak coupling logarithmic scheme that ratio is
temperature- and field-independent. Nevertheless, we found that, similar to Ref.\cite{LW}, the
gapless state is most visible when the superconducting gap on the heavier band is driven by the 
superconducting transition on the lighter band.

Strong anisotropy of the $B-T$ diagram, which indicates a quasi-2D nature of 2H-NbSe$_2$, 
and its multi-gap superconductivity\cite{yokoya,fletcher,mazin} makes a case for a
possible first order phase transition to a ground state of this type in low magnetic fields parallel to the plane in
this material\cite{BG2}. An unusual first order phase transition has been indeed observed in this 
material\cite{sologubenko} for thermal conductivity measurements in low magnetic field $H \sim 10 kOe \ll H_{c2}$ 
parallel to the basal plane. This first order phase transition is inconsistent with explanations 
involving vortex lattice melting\cite{sologubenko}.
The magnetic field at which this transition occurs is consistent with the value of the small energy 
gap $\sim 0.1-0.2 meV$   observed in the photoemission experiments\cite{yokoya}. 
The low-field first-order phase transition  in NbSe$_2$ was found to be strongly anisotropic,
and the hysteretic behavior of thermal conductivity disappears for certain field directions\cite{sologubenko},
a behavior not expected in a simple multiband model considered above, where a first order phase transition
occurs for all field directions in the basal plane. However, in the presence of CDW the CI symmetry
is broken\cite{bulaevskii}, which itself has been shown to lead to a strong in-plane anisotropy of the $B-T$
phase diagram\cite{BG}. Similar effects must be present for the low-field metamagnetic phase transition as well.
Gap anisotropy, g-factor anisotropy, and spin-orbit coupling tend to wipe out the first order line
on the $B-T$ phase diagram, and turn it in a smooth crossover\cite{BG2}.
We have recently found that the order of the phase transition in the presence of spin orbit interaction may, too, be 
dependent on direction. As a result of two different terms in the spin orbit interaction, the first
order phase transition becomes very anisotropic. It is present for some field directions and turns
into a smooth crossover for other field directions\cite{BG3}. 

Experimental observation of the new gapless state is subject to the usual difficulties associated with 
the observation of paramagnetic pair breaking and the LOFF state in superconductors. Namely, the orbital
effects leading to $H_{c2}$ are almost always present, even in strongly quasi-2D materials in magnetic
fields parallel to the 2D planes.  Perhaps, an ideal realization of this state would be surface 
superconductivity or superconductivity in thin films in fields parallel to the surface. 
We note, however, that unlike the LOFF state, which is difficult to observe, since it quickly disappears in the presence
of impurities or orbital effects, the new homogeneous partially gapless state is more robust, and
will appear in many multiband strongly quasi-2D s-wave superconductors in the mixed state as well, provided that
the upper critical field is close enough to the Clogston limit. The details of such first order transition in the
mixed state will be similar to the physics considered above.
Measurements of the specific heat in applied field are the most direct  way to observe the 
low-field first order phase transition in s-wave multiband superconductors,  such as 2H-NbSe$_2$\cite{yokoya}. 

\section{Acknowledgements}

The author is very grateful to Prof. L.P. Gor'kov for many discussions, and to  
Dr. A.V. Sologubenko for sharing new experimental results. This work was supported by TAML at the University of 
Tennessee.

\end{document}